\documentstyle{esub2acm}
\title{Partial Evaluation for Program Comprehension}
\author{Sandrine Blazy \and Philippe Facon }
\sponsor{CEDRIC-IIE, 18 all\'ee Jean Rostand, 91025 \'Evry Cedex, France
                     (blazy@iie.cnam.fr)}
\begin{document}
\begin{bottomstuff}
\permission
\end{bottomstuff}
\maketitle

\section{Introduction}
Program comprehension is the most tedious and time consuming task of software 
maintenance, an important phase of the software life cycle \cite{frazer:reverse}. 
This is particularly true while maintaining scientific application programs that 
have been written in Fortran for decades and that are still vital in various 
domains even though more modern languages are used to implement their user 
interfaces. Very often, programs have evolved as their application domains 
increase continually and have become very complex due to extensive modifications. 
This generality in programs is implemented by input variables whose value does 
not vary in the context of a given application. Thus, it is very interesting for 
the maintainer to propagate such information, that is to obtain a simplified 
program, which behaves like the initial one when used according to the restriction. 

We have adapted partial evaluation for program comprehension. Our partial 
evaluator performs mainly two tasks: constant propagation and statements simplification. 
It includes an interprocedural alias analysis. As our aim is program comprehension rather 
than optimization, there are two main differences with classical partial evaluation.
Firstly, we do not change the original structure of the code. In particular, we do not 
unfold statements. In the same way, our partial evaluator generates neither new variables 
nor rename variables. The residual code is easier to understand because many statements 
and variables have been removed and no additional statement or variable has been inserted.

Secondly, some identifiers are not replaced by their corresponding values in the residual 
code. The benefit of replacing an identifier by its value depends on the meaning of the 
identifier for the user: thus, it depends on the kind of identifier, but also on the kind of 
user. For any user, identifiers like PI are likely to be kept in the code, on the contrary 
to intermediate variables used only to decompose some computations. A physicist who is 
familiar with the equations implemented in the code may prefer to keep variables that are 
meaningful for him; on the contrary other users may prefer to see as few variables as 
possible. In fact, our partial evaluator is very flexible in that respect. Of course, 
even when there is no replacement, the known value of a variable is kept in the 
environment of our simplification task, as it can give opportunities to remove useless code.

\section{Formal development of the partial evaluator}
Our partial evaluator - as software maintenance tool - must introduce absolutely no 
unforeseen changes in programs. Therefore, we have used a formal development method. The 
partial evaluator's behavior is described in natural semantics \cite{kahn:natsem} augmented with 
various set operators: the simplification is inductively defined, by inference rules on 
the Fortran abstract syntax. 
These formal concepts were very useful to clarify concepts of Fortran (e.g. common blocks 
in Fortran 77, pointers in Fortran 90) and to model complex transformations. They also
allowed us to prove the correctness of the partial evaluation, with respect to the 
dynamic semantics of Fortran 90, also given in natural semantics \cite{tapsoft:sb}.
 Last, our specification is abstract enough to be easily adapted to any imperative language.

\section{Description of the tool}
The partial evaluator has been implemented on top of a kernel that has been generated by 
the generic programming environment Centaur \cite{Centaur}. When provided with the 
description of a particular programming language, including its syntax and semantics, 
Centaur produces a language specific environment. We have merged two specific 
environments into an environment for partial evaluation: a Fortran 90 environment,
and an environment dedicated to a language that we have defined for expressing the scope 
of general constraints on variables.

The formal specifications have been implemented in a language provided by Centaur, 
called Typol, intended to be an implementation of natural semantics. Thus, the Typol 
rules are close to the formal specification rules. Typol programs are compiled into 
Prolog code. Set operators have been written directly in Prolog. Although our partial 
evaluation propagates only equalities (and some specific inequalities), our initial 
constraints on input variables are written in a general language for expressing 
relations between variables and values, because our next work will be to progragate 
such relations.

As our partial evaluator is a program comprehension tool, we have implemented a 
sophisticated graphical interface to facilitate the exploration of Fortran application 
programs. It has been written in Lisp, enhanced with structures for programming 
communication between graphical objects and processes. Different windows visualize with 
hyperlinks specialized versions of a procedure, propagated data, initial and residual 
application programs. Usually initial application programs consist of several Fortran 
files and each file is a Fortran procedure with about 150 lines of statements. 
Furthermore several instances of the tool can be triggered in parallel. The first 
experiments with that tool at EDF (the French electricity provider) are very encouraging 
\cite{ase:sb}.

\section{Conclusion}
Partial evaluation appears to be a promising technique not only for program optimization 
but also for program comprehension by allowing to focus on a specific context of the
computation. Our partial evaluator may be used in two ways: by visual display of the 
simplified program as part of the initial program (for documentation or debugging), or 
by generating this simplified program as an independent (executable) program.

We are currently working on the propagation of more general constraints than equalities. 
Furthermore, partial evaluation is complementary to program slicing
\cite{slicing:gallagher}, another technique for extracting code when debugging a program. 
Program slicing aims at identifying the statements of a program which impact directly or
indirectly on some variables values. We believe that merging partial evaluation (a forward 

walk on the call graph) and program slicing (a backward walk) would improve a lot the
 reduction of programs.

\end{document}